\documentclass[conference]{IEEEtran}
\IEEEoverridecommandlockouts

\usepackage[T1]{fontenc}
\usepackage[utf8]{inputenc}
\usepackage{amsmath,amssymb}
\usepackage{graphicx}
\usepackage{booktabs}
\usepackage{array}
\usepackage{tabularx}
\usepackage{xcolor}
\usepackage{pgfplots}
\pgfplotsset{compat=1.17}
\usepackage{subcaption}
\usepackage{url}
\usepackage[hidelinks]{hyperref}

\graphicspath{{figures/}}

\begin{document}

\title{A Large Language Model (LLM) Approach to\\COVID-19 Fake News Detection}

\author{
\IEEEauthorblockN{Muhammad Arif\IEEEauthorrefmark{1},
Samah Fodeh-Jarad\IEEEauthorrefmark{1},
Elyas Irankhah\IEEEauthorrefmark{1},
Uroosa Fatima\IEEEauthorrefmark{2},
Lama Ayash\IEEEauthorrefmark{3}, and
Sreeraj Ramachandran\IEEEauthorrefmark{1}}
\IEEEauthorblockA{\IEEEauthorrefmark{1}Department of Emergency Medicine, Yale School of Medicine,\\ Yale University, New Haven, CT, USA}
\IEEEauthorblockA{\IEEEauthorrefmark{2}Instituto Polit\'ecnico Nacional, Mexico}
\IEEEauthorblockA{\IEEEauthorrefmark{3}King Khalid University, Saudi Arabia}
}

\maketitle

\begin{abstract}
Text classification is a widely researched area within the domains of information retrieval and data mining, offering valuable applications in various fields, such as healthcare diagnosis, targeted marketing, entertainment, and the classification of COVID-19-related fake news. This problem has gathered significant attention from researchers globally, particularly due to recent advancements in data mining and natural language processing. These advancements have prompted the development of automated systems to efficiently classify textual data. Natural Language Processing (NLP) enables the classification of documents containing diverse textual content. Social media platforms generate a vast volume of data through user interactions. The methodology of the current research involves gathering a dataset. Within this realm, datasets such as the COVID-19 English Tweet Dataset have been instrumental for experimental purposes. This dataset encompasses various forms of textual content, including tweets, posts, and messages pertaining to the coronavirus pandemic. In an effort to harness the potential of transfer learning (TL) and large language models (LLMs), the proposed research applies pretrained transformer-based models and LLMs, namely BERT-base, ALBERT-base, XLM-RoBERTa-base, DistilBERT, GPT-2, XLNet, and T5, to this dataset. Additionally, six different optimization techniques, namely weighted fusion, particle swarm optimization (PSO) fusion, the Nelder--Mead algorithm, Powell's method, the Broyden--Fletcher--Goldfarb--Shanno (BFGS) algorithm, and the trust-region constrained optimization algorithm, are applied to enhance the models' results. The results indicate that the GPT-2 large language model demonstrates high accuracy and efficiency. The model's performance is evaluated using accuracy, precision, recall, and F1-score, with a score of 0.96 achieved for each metric. Moreover, heatmaps are created to visualize the outcomes obtained from each model. Finally, the study highlights the value of applying LLMs, identifies the best performer among all the classifiers, and outlines future directions for further exploration in this domain.
\end{abstract}


\section{Introduction}
Natural Language Processing (NLP) is defined as a scientific discipline which aims at providing computers with the ability to understand and generate natural language. NLP has recently received further attention given the expansion of studies in the area and increased computational capabilities. There are now efforts to find meaningful data hidden in language as a useful tool in education, healthcare, industry, and retail. Early work employed rule-based methods to solve NLP problems, which gave way to machine learning (ML) methods due to the richness of human language. These methods, which include Support Vector Machine (SVM) and Na\"ive Bayes classifiers, are more effective in addressing the notions and dynamics of natural language.

Text mining is connected with the NLP concept and deals with obtaining valuable information from the text present on social media platforms. This type of extraction is achieved through algorithms and methods that work with textual data. Despite being an essential part of AI, NLP relies on advanced approaches such as text mining and data analysis. Data mining, an integration of machine learning and deep learning, is a versatile tool with high success rates in several areas. It improves question answering, anaphora resolution, bioinformatics, and web-based network analysis~\cite{b1}.

According to previous findings, NLP and data mining are widely used techniques in sentiment analysis~\cite{b2}, hate speech identification, and assessing Urdu speeches~\cite{b3} and movie reviews~\cite{b4}. Many of these technologies are global and have a well-known influence on different aspects of human life. Social media is one of the most popular digital platforms in recent years. These platforms are frequently used and transmit enormous quantities of data. However, they have also been associated with the rise of hate speech. For instance, when a new film is launched, individuals make many posts on social media about it, whether good or bad. Hate speech has received a lot of attention from scholars due to the increase in hate speech incidents. A notable paper~\cite{b5} clarified the pivotal role of NLP in hate speech detection, showcasing how automated processes can effectively identify and flag instances of hatred within social media content.

Understanding and processing human-constructed natural language strictly demands NLP, which provides fundamental tools. The information produced by users on social media and review platforms contains useful knowledge for various sectors such as healthcare, industry, science, and policy-making. This content is rich in text and has the potential to inform decision-making in diverse areas. Leveraging opinion extraction algorithms and sentiment analysis techniques~\cite{b6}, researchers can extract valuable insights from this data. Notably, these algorithms play a crucial role in enhancing the performance of text classification models during the feature extraction phase~\cite{b7,b8,b9}. This approach helps stakeholders understand consumer sentiment, market trends, and public opinion.

In their work, a team of scholars described in detail the current shortcomings and loopholes in the field of hate speech detection~\cite{b10}. They presented a solution that aimed at overcoming these challenges, focusing on the lack of ample data needed to train accurate hate speech detection algorithms. The lack of adequate labeled datasets is a major challenge in accurately detecting hate speech occurrences on social media. To overcome this challenge, the researchers encouraged the use of pre-trained models like BERT since it is effective in contextual text analysis~\cite{b11}. Their approach is sensitive to changes in data over time and is exemplified by the AlBERTo model trained on an Italian hate speech dataset. By using only the latter half of the data for training the models, their accuracy significantly increases. The study further establishes the differences between hate speech and offensive language. Their method improves text classification by using users' entries on social media platforms and metaheuristic algorithms; therefore, it can be considered a promising approach to counter hate speech on social media platforms.

Transfer learning is a highly effective technique which utilizes information from unlabeled data to enhance the overall performance on tasks that have a small number of labeled samples. Prior information makes it possible to obtain high accuracy even with a small number of labeled data points. This is especially the case in NLP, where transformers outcompete traditional ML and DL techniques. Transfer learning obtains knowledge from related domains so that ML systems adapt to new tasks with higher accuracy. Thus, it minimizes the demand for human intervention and performs better than active and supervised learning. For example, a system trained to recognize apples can be tweaked to recognize pears, cutting down on the data needed and the time taken to train the system.

Social media platforms wield significant influence and provide users with a platform to freely express their thoughts and opinions. Leveraging this, researchers have used advanced machine learning models to distinguish between real and fabricated information, especially in relation to news about COVID-19. Among these models, transformers have proven to be particularly effective, providing superior performance in distinguishing between real and fake content.

News dissemination is of enormous importance and serves as a crucial source of information that keeps people informed and engaged. However, the spread of fake news dates back to the 18th century~\cite{b12}, and the advent of the Internet has exacerbated its spread, primarily through the widespread use of social media channels. Detecting the authenticity of news articles poses a huge challenge, and existing approaches show limitations. Researchers are developing hybrid methods to improve the accuracy and consistency of fake news detection and overcome current limitations.

In this work, we use various optimization techniques to improve model performance, reduce losses, and improve overall accuracy. Using state-of-the-art algorithms, the model parameters are refined to achieve optimal results, ensuring the robustness and reliability of the predictions. Additionally, ChatGPT was used as a writing and analysis aid during our experiments and the preparation of this paper, to help refine methodologies, interpret results, and articulate findings.

The remainder of the paper is organized as follows: Section~\ref{sec:lit} reviews the relevant literature, Section~\ref{sec:methods} describes the dataset, methodology, and classification algorithms, Section~\ref{sec:results} presents the experimental results, Section~\ref{sec:discussion} discusses them, and Section~\ref{sec:conclusion} presents the conclusions and future directions.

\section{Relevant Literature}\label{sec:lit}
In modern research, transfer learning stands as a fundamental technique that strives for greater precision in various domains. Researchers explore different transformer variants to advance their investigations effectively. In a recent study, researchers examined the latest advances in TL within the field of NLP~\cite{b13}. Their comprehensive analysis involved examining new ML and DL methodologies, as well as exploring modern TL approaches.

The researchers observed that TL techniques have introduced innovative dimensions in various NLP tasks, particularly benefiting scenarios with limited training data. By leveraging pre-trained models and subsequent fine-tuning, TL models demonstrate superior performance over earlier state-of-the-art NLP methods. Pre-trained and fine-tuned models act as a powerful transfer mechanism, but their efficiency can vary from task to task, prompting the development of new techniques to address complex challenges~\cite{b14}.

As an alternative, researchers introduced adapter modules~\cite{b14}. These modules facilitate the creation of scalable models, where each added task requires a minimal number of trainable parameters, avoiding the need to revisit previous tasks. Because the original model parameters remain fixed, stability increases. Evaluating BERT on 26 different classification tasks, including the GLUE benchmark, adapters attained near state-of-the-art performance while adding only 3.6\% additional parameters per task, compared with fine-tuning all parameters.

Recent research shows the effects of misinformation about COVID-19~\cite{b15}. It highlights how social media contained many unreliable messages that caused widespread panic. The authors present a strong approach to detecting fake news based on CT-BERT (COVID-Twitter-BERT), which achieved an F1-score of 98.69.

Researchers also developed a method to check whether social media posts about COVID-19 were credible~\cite{b16}. Their approach used TL models such as BERT, ALBERT, and XLNet. When tested on a large dataset, it achieved an F1-score of 98.55 and ranked 5th out of 160 teams. By carefully dividing the dataset, the study created a solid evaluation protocol, highlighting the importance of disseminating reliable information during crises.

Social media makes finding news very easy nowadays, but unfortunately fake news also spreads rapidly. To combat this, researchers studied features such as the actual content, who shared it, and their profiles, using these clues to detect fakes~\cite{b17}. However, this approach requires reliable labeled datasets, which are difficult to obtain. Therefore, an Unsupervised Fake News Detection technique (UFD) was developed. UFD was tested using 12,800 short statements and 1,627 American election articles, and outperformed other unsupervised methods in precision, recall, F1-score, and accuracy, with accuracy of 0.759 on LIAR and 0.679 on BuzzFeed data.

In an analysis evaluating strategies for recognizing fake news during the COVID-19 pandemic, researchers explored deep learning and transfer learning strategies~\cite{b18}. Analyzing a dataset of 10,700 social media posts and articles, they employed several algorithms, including Bi-LSTM with attention, the Hierarchical Attention Network (HAN), DistilBERT, and BERT-base, to distinguish between real and fake news. They focused exclusively on news content, ignoring user-level or propagation signals from social media, which are not always available. Interestingly, further pre-training BERT and DistilBERT on COVID-19 tweets provided superior overall performance compared to models fine-tuned only on the dataset. In particular, the BERT model pre-trained in this way performed better than the COVID-Twitter-BERT model.

In~\cite{b19}, researchers examined the prevalent effect of social media on daily life, shedding light on its potential dangers. They addressed the problem of misinformation, especially COVID-19 news. Their approach to distinguishing real information from fake earned a high F1-score and secured second place in the rankings. Using datasets drawn from platforms such as Facebook, Twitter, and Instagram, divided into training, validation, and test sets, and experimenting with various baseline and transformer models, the researchers validated the superiority of their RoBERTa and ELECTRA models, achieving F1-scores of 0.9864 and 0.9827, respectively, on the official test set.

Another study proposed an approach to political text classification that leverages the insights of human experts through annotation of political manifestos~\cite{b20}. This annotated corpus serves as training data for a classifier, which is then applied to COVID-19 press conferences. By combining convolutional neural networks (CNN) with BERT, the researchers achieved strong results. Among the models examined (M1--M4), M4, which combines CNN and BERT, was the most effective, with the highest accuracy and F1-score.

In an investigation of sentiment analysis in finance, researchers tested recent improvements in the field~\cite{b21}. Their research encompassed diverse methodologies, from traditional lexicon-based approaches to modern transformer models. Among these, transformers emerged as the best performers. The group evaluated an array of transformer models, including BERT, FinBERT, and XLNet, among others. Notably, BART stood out with an F1-score of 0.85, showcasing its ability to capture complex financial sentiment.

BERT, trained on large corpora such as BookCorpus and Wikipedia, achieves strong results in many NLP tasks~\cite{b22}. However, its efficacy diminishes in certain domains and tasks, revealing room for improvement. Addressing this gap, researchers introduced BERT4TC, a variant tailored for text classification that incorporates domain-specific and task-oriented knowledge. Evaluated on publicly available datasets, BERT4TC outperformed conventional feature-based approaches and several fine-tuned methods, establishing new benchmarks in multiclass classification.

The Internet has been seamlessly integrated into our lives and has become an essential part of our routines~\cite{b23}. While it serves as a trusted source of knowledge and information, it has also highlighted concerning issues. The prevalence of harmful language online underscores the importance of taking action, and advances have been made in this area through ML techniques. An example is the DeepHate model, which was developed to efficiently analyze such data and detect patterns of hate speech. With transfer learning as its basis, the model captures not only word meanings but also sentence structure. Studies on datasets such as English tweets (37,520 tweets) represent a step forward in the fight against online harassment.

\section{Material and Methods}\label{sec:methods}
\begin{figure*}[t]
\centering
\includegraphics[width=0.85\textwidth]{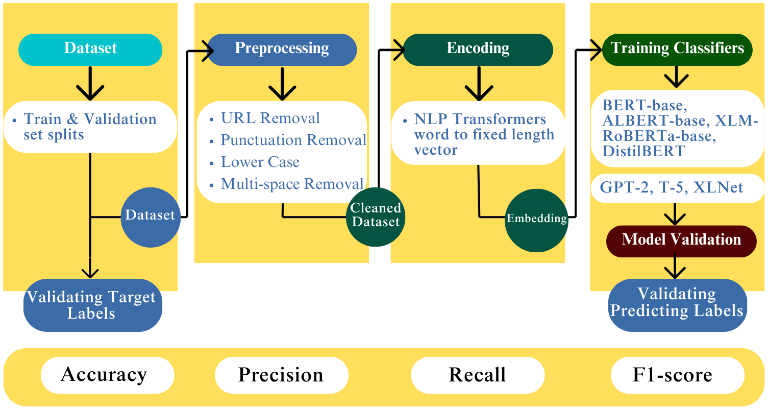}
\caption{Block diagram of the proposed methodology.}
\label{fig:block}
\end{figure*}

The overall methodology used to classify COVID-19 news as real or fake is shown in Fig.~\ref{fig:block}. It includes dataset collection, data preprocessing, data splitting, and model evaluation.

\subsection{Dataset}
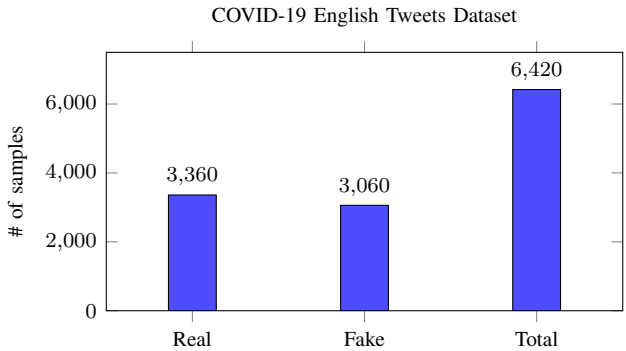
\begin{figure}[t]
\centering
\begin{tikzpicture}
\begin{axis}[
  ybar, bar width=18pt, width=0.95\columnwidth, height=5cm,
  ymin=0, ymax=7500, ylabel={\# of samples},
  symbolic x coords={Real,Fake,Total}, xtick=data,
  nodes near coords, nodes near coords style={font=\footnotesize},
  title={COVID-19 English Tweets Dataset},
  title style={font=\footnotesize}, tick label style={font=\footnotesize},
  label style={font=\footnotesize}, enlarge x limits=0.25]
\addplot[fill=blue!70] coordinates {(Real,3360) (Fake,3060) (Total,6420)};
\end{axis}
\end{tikzpicture}
\caption{Structure of the dataset.}
\label{fig:dataset}
\end{figure}

This study analyzed the ``COVID-19 English Tweets'' dataset~\cite{b55}. The dataset is labeled with 3,360 real tweets and 3,060 fake tweets, a total of 6,420 tweets (Fig.~\ref{fig:dataset}), and offers valuable insights into pandemic-related social media discourse.

The dataset is publicly available and contains two columns. The first column, ``Tweets'', contains the textual content, while the second column, ``Labels'', distinguishes authentic tweets from fake ones. Examples are presented in Table~\ref{tab:examples}.
\begin{itemize}
\item \textbf{Tweets:} This column contains the original textual data representing tweets related to the COVID-19 pandemic. Each row contains the textual content of a particular tweet, ranging from updates, statistics, and precautions to opinions concerning the pandemic.
\item \textbf{Labels:} This column specifies the authenticity of the information contained in each tweet. Tweets are categorized as ``real'' when they contain accurate information and as ``fake'' when they contain potentially misleading information related to the pandemic.
\end{itemize}

\begin{figure}[t]
\centering
\includegraphics[width=\columnwidth]{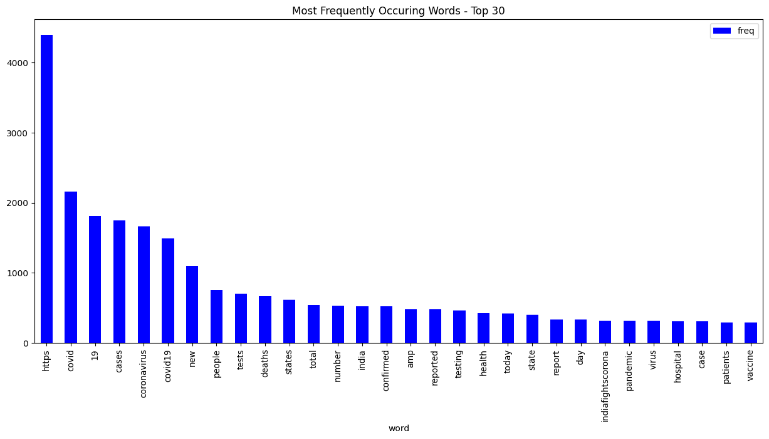}
\caption{Frequency of the top 30 most frequent words in the raw data.}
\label{fig:freqraw}
\end{figure}

\begin{figure}[t]
\centering
\includegraphics[width=0.9\columnwidth]{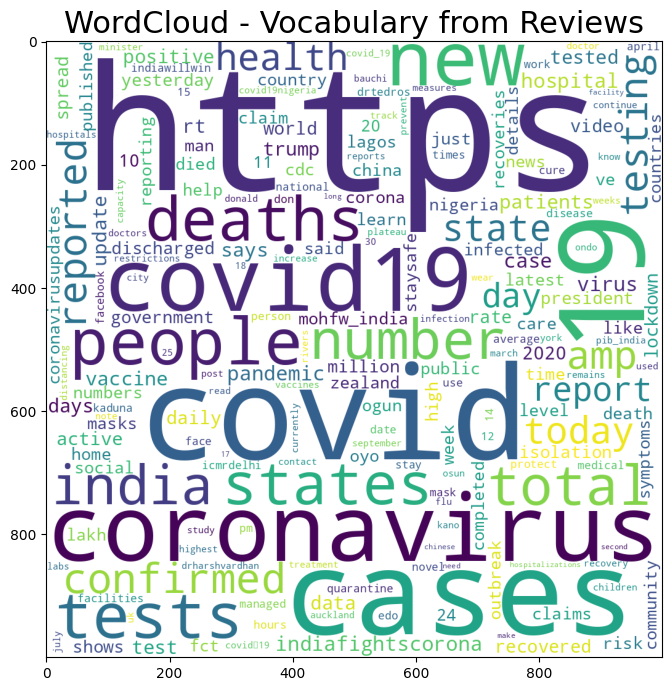}
\caption{Word cloud of the raw data.}
\label{fig:wcraw}
\end{figure}

To process the data, real labels were replaced with 1 and fake labels with 0 using a Python script. Fig.~\ref{fig:freqraw} illustrates the frequency of the 30 most commonly occurring words, and Fig.~\ref{fig:wcraw} shows a word cloud generated from the raw, unprocessed data.

\begin{table*}[t]
\centering
\caption{Example tweets from the COVID-19 English Tweets dataset}
\label{tab:examples}
\begin{tabularx}{\textwidth}{X c}
\toprule
\textbf{Tweet} & \textbf{Label}\\
\midrule
The NBA is poised to restart this month. In March we reported on how the Utah Jazz got 58 coronavirus tests in a matter of hours at a time when U.S.\ testing was sluggish. \url{https://t.co/I8YjjrNoTh} \url{https://t.co/o0Nk6gpyos} & Fake\\
\midrule
We just announced that the first participants in each age cohort have been dosed in the Phase 2 study of our mRNA vaccine (mRNA-1273) against novel coronavirus. Read more: \url{https://t.co/woPlKz1bZC} \#mRNA \url{https://t.co/9VGUoJu5cS} & Real\\
\bottomrule
\end{tabularx}
\end{table*}

\subsection{Data Preprocessing}
\begin{figure}[t]
\centering
\includegraphics[width=\columnwidth]{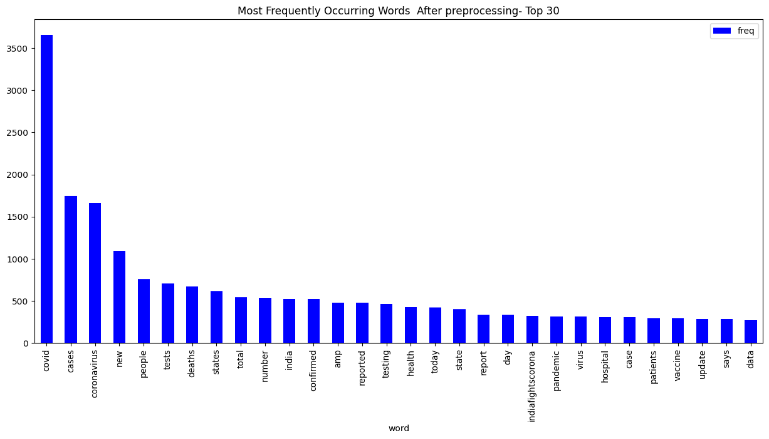}
\caption{Frequency of the top 30 most frequent words in the processed data.}
\label{fig:freqproc}
\end{figure}

\begin{figure}[t]
\centering
\includegraphics[width=0.9\columnwidth]{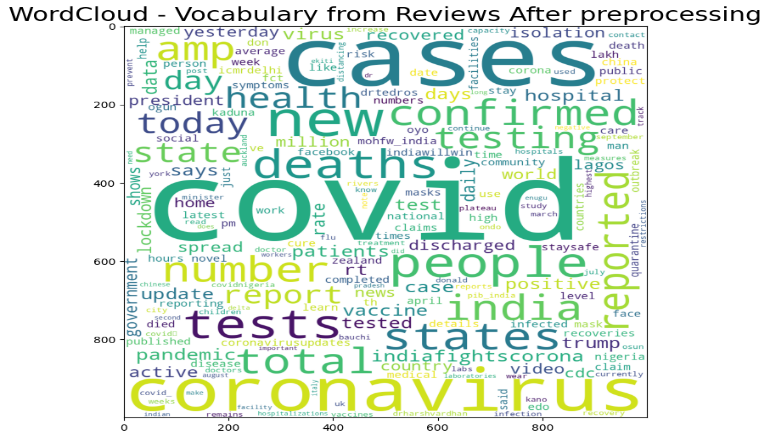}
\caption{Word cloud of the processed data.}
\label{fig:wcproc}
\end{figure}

After collecting and organizing the dataset, the next step was data preprocessing, which focused primarily on data cleaning: removing URLs, special tags, and numbers, and replacing multiple spaces with a single space. Additionally, all characters were converted to lowercase for consistency. Fig.~\ref{fig:freqproc} and Fig.~\ref{fig:wcproc} show the 30 most frequent words and the word cloud of the cleaned and preprocessed dataset. No class-balancing approach was applied, as the dataset contains enough samples of both classes for the experiments.

\subsection{Training, Validation, and Test Splits}
After cleaning and preprocessing, the dataset was divided into training, validation, and test sets containing approximately 70\%, 10\%, and 20\% of the data, respectively. Models are therefore trained on 70\% of the data, validated on 10\%, and tested on the remaining 20\%.

\subsection{Model Evaluation}
After training, model performance was evaluated using precision, recall, accuracy, and F1-score. Confusion matrices, visualized as heatmaps, were used to understand how well each model performs by comparing predicted values with actual values. The confusion matrix is a simple $M \times N$ matrix containing true positives, true negatives, false positives, and false negatives~\cite{b24}; it shows where the model excels and where it needs improvement.

\subsection{Classification Algorithms}
This study explored seven classification models: BERT-base, XLM-RoBERTa-base, ALBERT-base, and DistilBERT for TL, along with three LLMs, namely GPT-2, T5, and XLNet. LLMs are advanced machine learning models designed to understand, generate, and manipulate human language. They are typically built using deep learning techniques and trained on vast amounts of textual data.

Fine-tuned models were used in this study. The parameters of the selected TL models and LLMs were initialized with their default settings, except for some that were altered as needed, such as the number of epochs, learning rate, and batch size. For example, the default learning rate for GPT-2 is $5 \times 10^{-5}$, whereas in this study it was set to $2 \times 10^{-5}$.

First, the TL models were evaluated on the dataset to assess their performance. Optimization techniques were then applied to further improve their results. By comparing model effectiveness before and after optimization, the study aimed to achieve better results.

\subsubsection{BERT-Base}
BERT, introduced in~\cite{b25}, is an abbreviation for Bidirectional Encoder Representations from Transformers. BERT is pre-trained on unlabeled text and takes into account context at all levels, making it a powerful tool in natural language processing. Adapting a BERT model for a task is simple: one adds a task-specific output layer. BERT uses a Masked Language Model (MLM) objective, in which words are hidden in the input text and the model predicts them based on the surrounding words. By considering both left and right contexts, bidirectional understanding is learned. Additionally, it combines MLM with Next Sentence Prediction (NSP). BERT-Base requires comparatively less processing and computation time. It performs well in tasks such as fake news detection, content refinement, and knowledge distillation, but is less suited to fuzzy data or text mining~\cite{b24}. BERT has been used to detect fake news~\cite{b26}, for content-enhanced text-to-SQL generation~\cite{b27}, and for sentiment analysis~\cite{b28,b29}.

\subsubsection{ALBERT-Base}
In 2019, Toyota Technological Institute at Chicago and Google Research introduced ALBERT~\cite{b30}, a compact successor to BERT. By reducing parameters, ALBERT accelerates training and reduces memory usage. ALBERT performs well in various classification tasks and in sentiment analysis despite its small parameter count~\cite{b24}. ALBERT was employed to detect sarcasm on Twitter~\cite{b31}, applied to fake news detection where it outperformed other models~\cite{b32}, evaluated for COVID-19 question answering with an exact-match score of 13.04~\cite{b33}, and used to analyze medication use reported on social media~\cite{b34}.

\subsubsection{DistilBERT}
In 2019, DistilBERT~\cite{b35} was introduced as a smaller, faster, and cheaper variant of BERT, being 60\% faster and 40\% smaller while retaining 97\% of its language understanding capability~\cite{b24}. The authors of~\cite{b36} examined the ability of DistilBERT to classify socio-political news, with promising results. Another study~\cite{b37} analyzed compositional and lexical semantics in RoBERTa, BERT, and DistilBERT. DistilBERT has also been used to detect protected health information and medical concepts through named entity recognition~\cite{b38}, enhanced for answer selection with a 0.6\% improvement~\cite{b39}, and retrained on universal dependencies for a voice shopping assistant~\cite{b40}.

\subsubsection{XLM-RoBERTa-Base}
XLM-RoBERTa builds on RoBERTa~\cite{b41} and is trained across one hundred languages using over two terabytes of filtered data. In~\cite{b46}, researchers explored this model for multilingual sentiment analysis, reporting a good F1-score. For EVALITA 2020, researchers in~\cite{b47} achieved an F1-score of 0.798 on a hate speech dataset, and~\cite{b48} demonstrated the superior performance of XLM-RoBERTa in word-in-context disambiguation. The authors of~\cite{b45} combined XLM-RoBERTa with DPCNN for offensive language identification. Related work has applied RoBERTa to linguistic generalization~\cite{b42}, mental illness detection~\cite{b43}, and informative COVID-19 tweet detection~\cite{b44}.

\subsubsection{GPT-2}
GPT-2 (Generative Pre-trained Transformer 2) is an autoregressive language model that generates text token by token, capturing context to produce fluent text across many topics~\cite{b49}.

\subsubsection{XLNet}
XLNet, developed by Google AI and Carnegie Mellon University, builds on Transformer-XL as its core architecture. Unlike BERT, which masks input tokens and does not fully capture dependencies between masked positions, XLNet learns bidirectional contexts by maximizing the expected likelihood over permutations of the factorization order, addressing these limitations of BERT~\cite{b58}.

\subsubsection{T5}
The Text-to-Text Transfer Transformer (T5), created by Google AI, treats every task as a text-to-text problem. T5 uses an encoder--decoder transformer pre-trained with an unsupervised span-corruption objective. While T5 has been used effectively for many NLP tasks, methods for extracting high-quality text representations from it are still being studied~\cite{b59}.

\subsection{Optimization Techniques and Fusion}
Various optimization techniques were explored to boost model performance. Following the fusion approach of~\cite{b50}, we use merit-based late fusion, combining classification scores from different models using various weight-selection and optimization methods. A separate validation set is used to calculate the classification error that serves as the fitness function for each method. These methods were selected based on their previous success in similar tasks~\cite{b51}.

\subsubsection{Weighted Fusion}
Weighted fusion improves prediction accuracy by giving more weight to better-performing models. Models with greater precision or reliability receive higher weights, so the final result reflects the strength of each model. Averaging predictions with these weights balances the influence of each model, resulting in a more reliable overall result.

\subsubsection{PSO Fusion}
Particle Swarm Optimization (PSO)~\cite{b52} is a widely used optimization technique known for its effectiveness and versatility. It has been applied to hyper-parameter tuning in neural networks~\cite{b54} and federated learning~\cite{b53}. PSO starts with a population of diverse candidate solutions and improves them iteratively based on a fitness function, updating particle positions and velocities using personal-best and global-best positions~\cite{b50}.

\subsubsection{Nelder--Mead Algorithm}
The Nelder--Mead algorithm, introduced in 1965, is one of the oldest and most widely used derivative-free optimization methods and is supported in numerous statistical and machine learning libraries~\cite{b51}. It is well suited to $n$-dimensional problems and to finding the minimum of an objective function, such as minimizing the classification error on validation data to obtain fusion weights~\cite{b50}.

\subsubsection{Powell's Method}
Powell's method is a popular derivative-free search method that performs successive line searches along a set of conjugate directions. Like PSO, it iteratively refines an initial guess, moving towards a minimum of the function. Despite its effectiveness, it has limitations, such as assumptions about step size, but it offers a simple yet effective optimization strategy~\cite{b50}.

\subsubsection{BFGS}
The Broyden--Fletcher--Goldfarb--Shanno (BFGS) algorithm is an iterative quasi-Newton method for minimizing functions in multidimensional spaces. It is useful when the gradient of the function is available and performs well on smooth, well-behaved problems, making it a common choice for unconstrained optimization.

\subsubsection{Trust-Region Constrained (trust-constr) Algorithm}
The trust-region constrained algorithm minimizes or maximizes a function while respecting constraints. It handles both equality and inequality constraints and uses trust regions for efficient optimization, making it well suited to constrained problems with smooth objective functions.

\section{Experiments and Results}\label{sec:results}
This section presents the results of each experiment on the dataset in terms of all performance metrics, together with the confusion-matrix heatmaps.

Table~\ref{tab:tl} shows the results of the four TL classifiers without optimization, Table~\ref{tab:opt} presents the results of the optimization techniques, and Table~\ref{tab:llm} reports the results of the three LLMs. Table~\ref{tab:sota} compares the proposed work with state-of-the-art approaches.

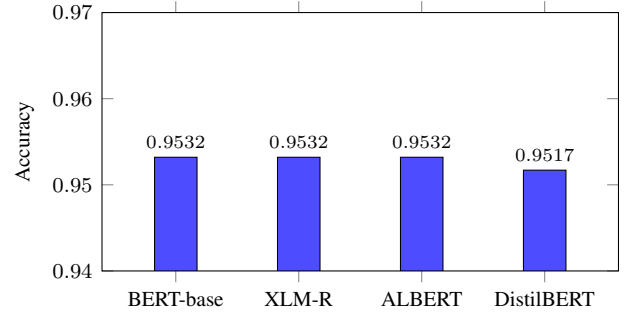
\begin{figure}[t]
\centering
\begin{tikzpicture}
\begin{axis}[
  ybar, bar width=16pt, width=0.95\columnwidth, height=5cm,
  ymin=0.94, ymax=0.97, ylabel={Accuracy},
  symbolic x coords={BERT-base,XLM-R,ALBERT,DistilBERT}, xtick=data,
  nodes near coords, nodes near coords style={font=\scriptsize, /pgf/number format/precision=4},
  tick label style={font=\footnotesize}, label style={font=\footnotesize},
  yticklabel style={/pgf/number format/fixed, /pgf/number format/precision=2},
  enlarge x limits=0.2]
\addplot[fill=blue!70] coordinates {(BERT-base,0.9532) (XLM-R,0.9532) (ALBERT,0.9532) (DistilBERT,0.9517)};
\end{axis}
\end{tikzpicture}
\caption{Accuracies of the transfer learning classifiers.}
\label{fig:tlacc}
\end{figure}

\begin{figure}[t]
\centering
\begin{tikzpicture}
\begin{axis}[
  ybar, bar width=12pt, width=0.98\columnwidth, height=5cm,
  ymin=0.94, ymax=0.97, ylabel={Accuracy},
  symbolic x coords={Weighted,PSO,Nelder-Mead,Powell,BFGS,Trust}, xtick=data,
  tick label style={font=\scriptsize}, label style={font=\footnotesize},
  yticklabel style={/pgf/number format/fixed, /pgf/number format/precision=2},
  enlarge x limits=0.1]
\addplot[fill=blue!70] coordinates {(Weighted,0.9595) (PSO,0.9563) (Nelder-Mead,0.9610) (Powell,0.9532) (BFGS,0.9532) (Trust,0.9532)};
\end{axis}
\end{tikzpicture}
\caption{Accuracies obtained with the optimization techniques.}
\label{fig:optacc}
\end{figure}
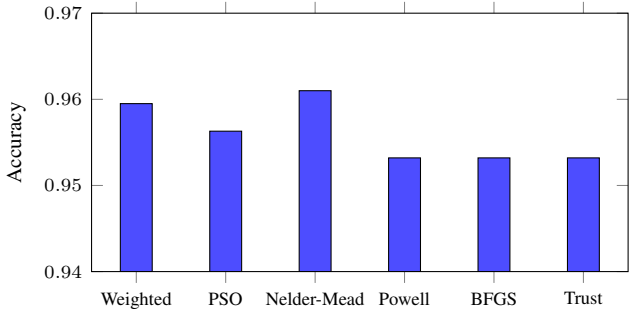

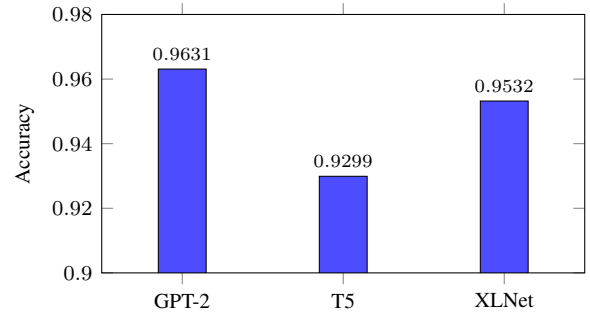
\begin{figure}[t]
\centering
\begin{tikzpicture}
\begin{axis}[
  ybar, bar width=18pt, width=0.9\columnwidth, height=5cm,
  ymin=0.90, ymax=0.98, ylabel={Accuracy},
  symbolic x coords={GPT-2,T5,XLNet}, xtick=data,
  nodes near coords, nodes near coords style={font=\scriptsize, /pgf/number format/precision=4},
  tick label style={font=\footnotesize}, label style={font=\footnotesize},
  yticklabel style={/pgf/number format/fixed, /pgf/number format/precision=2},
  enlarge x limits=0.25]
\addplot[fill=blue!70] coordinates {(GPT-2,0.9631) (T5,0.9299) (XLNet,0.9532)};
\end{axis}
\end{tikzpicture}
\caption{Accuracies of the large language models.}
\label{fig:llmacc}
\end{figure}

\begin{figure}[t]
\centering
\begin{tikzpicture}
\begin{axis}[
  ybar, bar width=8pt, width=0.95\columnwidth, height=5cm,
  ymin=0.90, ymax=0.98, ylabel={Score},
  symbolic x coords={GPT-2,T5,XLNet}, xtick=data,
  tick label style={font=\footnotesize}, label style={font=\footnotesize},
  yticklabel style={/pgf/number format/fixed, /pgf/number format/precision=2},
  legend style={font=\scriptsize, at={(0.5,-0.2)}, anchor=north, legend columns=3},
  enlarge x limits=0.25]
\addplot[fill=blue!70] coordinates {(GPT-2,0.9612) (T5,0.9356) (XLNet,0.9540)};
\addplot[fill=orange!70] coordinates {(GPT-2,0.9610) (T5,0.9314) (XLNet,0.9568)};
\addplot[fill=teal!70] coordinates {(GPT-2,0.9600) (T5,0.9310) (XLNet,0.9554)};
\legend{Precision,Recall,F1-score}
\end{axis}
\end{tikzpicture}
\caption{Precision, recall, and F1-score of the large language models.}
\label{fig:llmmetrics}
\end{figure}
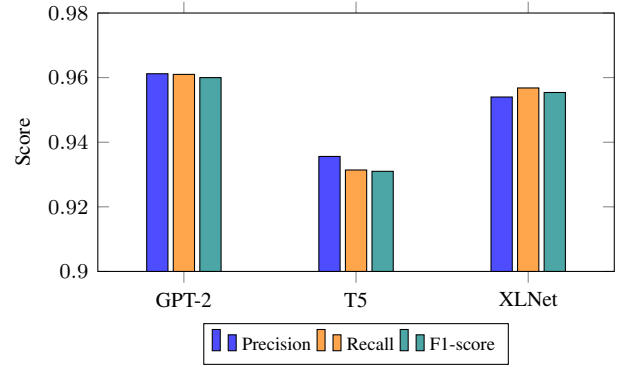

\begin{figure*}[t]
\centering
\begin{subfigure}{0.24\textwidth}\includegraphics[width=\linewidth]{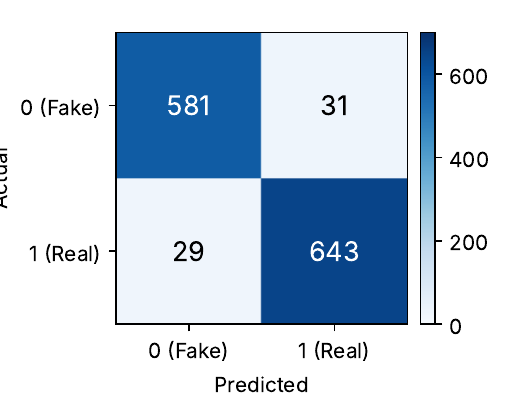}\caption{BERT-base}\end{subfigure}\hfill
\begin{subfigure}{0.24\textwidth}\includegraphics[width=\linewidth]{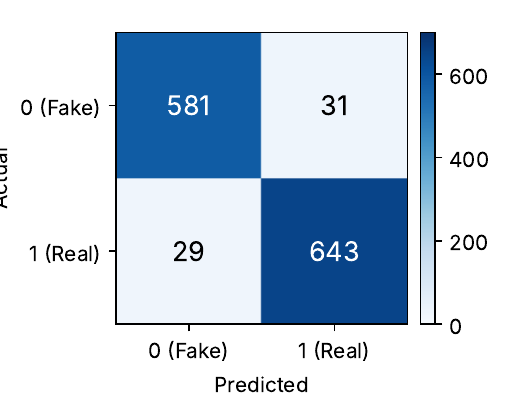}\caption{XLM-RoBERTa}\end{subfigure}\hfill
\begin{subfigure}{0.24\textwidth}\includegraphics[width=\linewidth]{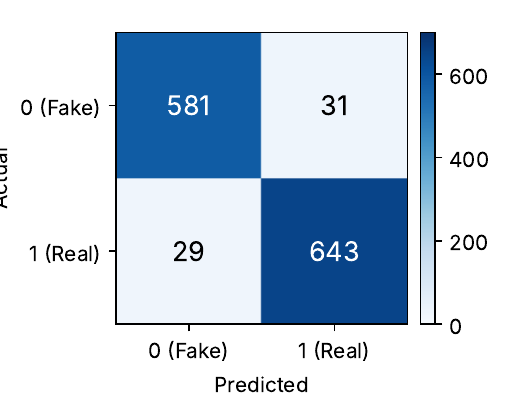}\caption{ALBERT}\end{subfigure}\hfill
\begin{subfigure}{0.24\textwidth}\includegraphics[width=\linewidth]{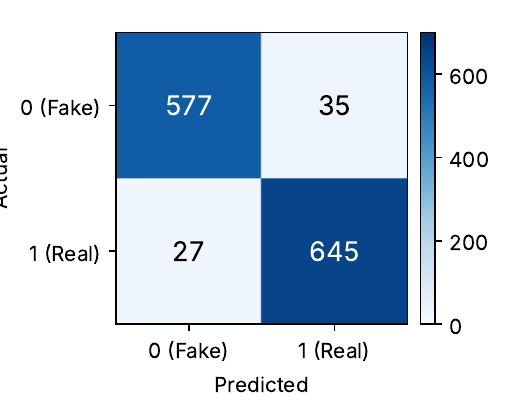}\caption{DistilBERT}\end{subfigure}
\caption{Confusion-matrix heatmaps of the transfer learning models.}
\label{fig:cmtl}
\end{figure*}

\begin{figure*}[t]
\centering
\begin{subfigure}{0.32\textwidth}\includegraphics[width=\linewidth]{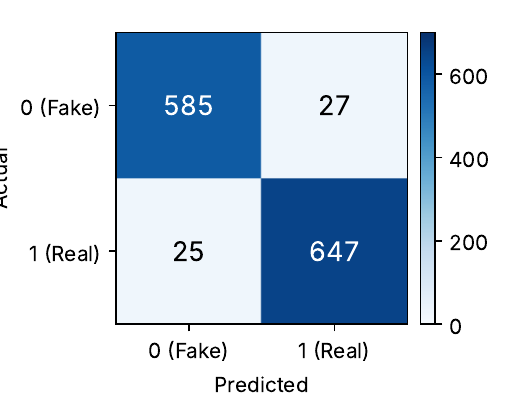}\caption{Weighted fusion}\end{subfigure}\hfill
\begin{subfigure}{0.32\textwidth}\includegraphics[width=\linewidth]{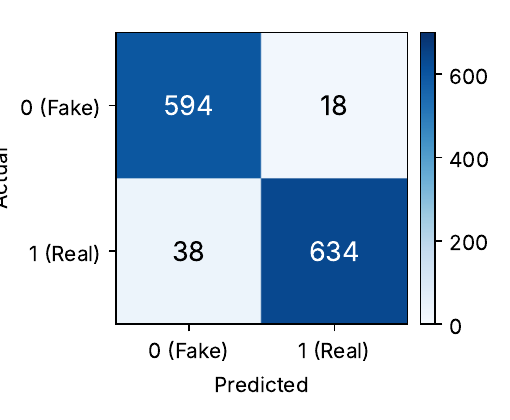}\caption{PSO fusion}\end{subfigure}\hfill
\begin{subfigure}{0.32\textwidth}\includegraphics[width=\linewidth]{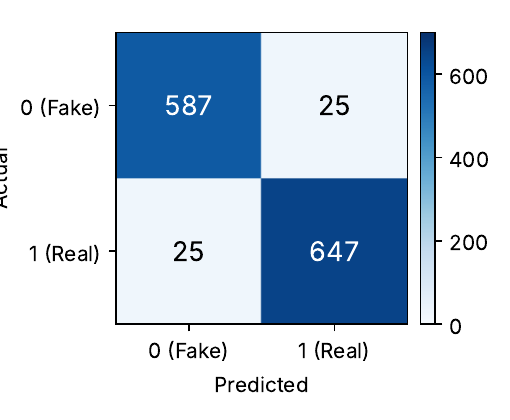}\caption{Nelder--Mead fusion}\end{subfigure}\\[4pt]
\begin{subfigure}{0.32\textwidth}\includegraphics[width=\linewidth]{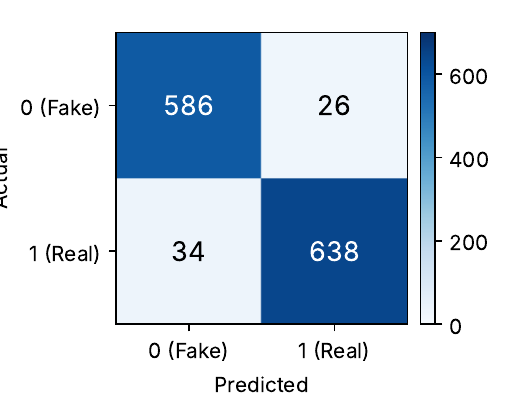}\caption{Powell fusion}\end{subfigure}\hfill
\begin{subfigure}{0.32\textwidth}\includegraphics[width=\linewidth]{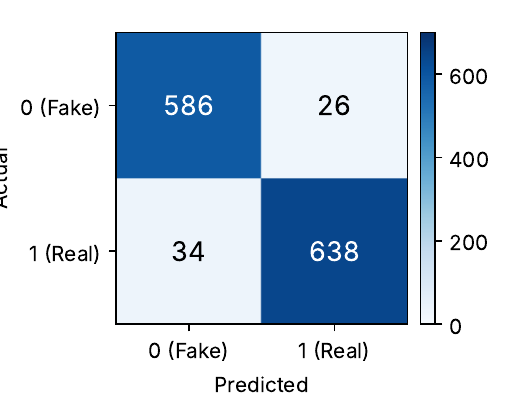}\caption{BFGS fusion}\end{subfigure}\hfill
\begin{subfigure}{0.32\textwidth}\includegraphics[width=\linewidth]{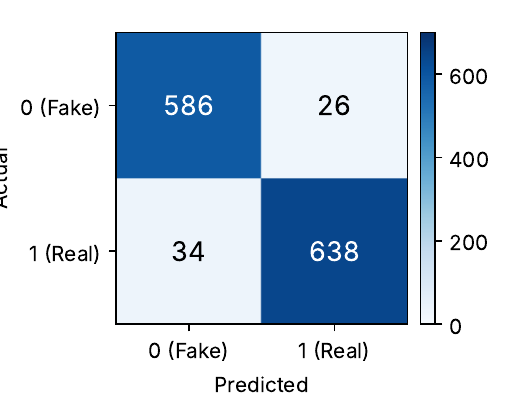}\caption{Trust-constr fusion}\end{subfigure}
\caption{Confusion-matrix heatmaps of the fusion/optimization techniques.}
\label{fig:cmopt}
\end{figure*}

\begin{figure*}[t]
\centering
\begin{subfigure}{0.32\textwidth}\includegraphics[width=\linewidth]{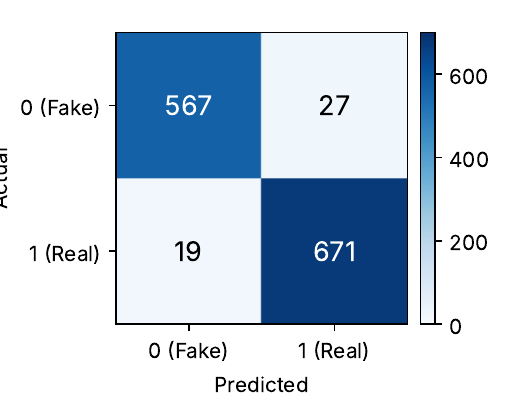}\caption{GPT-2}\end{subfigure}\hfill
\begin{subfigure}{0.32\textwidth}\includegraphics[width=\linewidth]{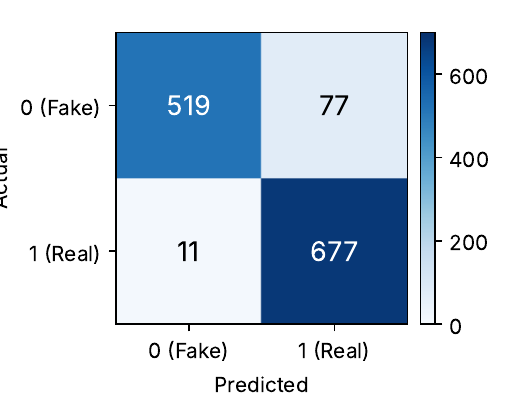}\caption{T5}\end{subfigure}\hfill
\begin{subfigure}{0.32\textwidth}\includegraphics[width=\linewidth]{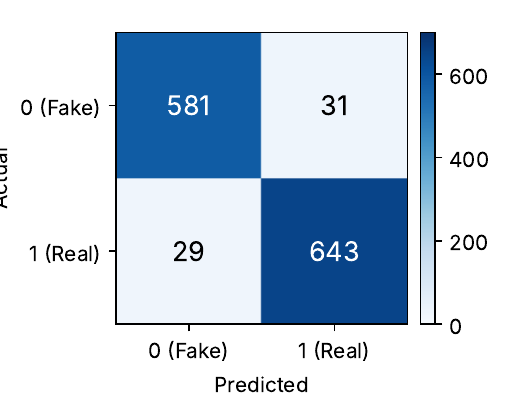}\caption{XLNet}\end{subfigure}
\caption{Confusion-matrix heatmaps of the large language models.}
\label{fig:cmllm}
\end{figure*}

Fig.~\ref{fig:tlacc} shows the accuracies of all TL models, and Fig.~\ref{fig:tlmetrics} presents their precision, recall, and F1-score. Fig.~\ref{fig:optacc} shows the results of the optimization techniques. Fig.~\ref{fig:llmacc} shows the accuracies of the LLMs and Fig.~\ref{fig:llmmetrics} their remaining metrics. Fig.~\ref{fig:cmtl} shows the heatmaps of the TL models, Fig.~\ref{fig:cmopt} those of the fusion/optimization techniques, and Fig.~\ref{fig:cmllm} those of the LLMs.

\begin{table}[t]
\centering
\caption{Results of the transfer learning models}
\label{tab:tl}
\begin{tabular}{lcccc}
\toprule
Model & Accuracy & Precision & Recall & F1\\
\midrule
BERT-base & 0.9532 & 0.9540 & 0.9568 & 0.9554\\
XLM-RoBERTa-base & 0.9532 & 0.9540 & 0.9568 & 0.9554\\
ALBERT-base & 0.9532 & 0.9540 & 0.9568 & 0.9554\\
DistilBERT & 0.9517 & 0.9485 & 0.9598 & 0.9541\\
\bottomrule
\end{tabular}
\end{table}

\begin{table}[t]
\centering
\caption{Results of the optimization techniques}
\label{tab:opt}
\begin{tabular}{lcccc}
\toprule
Method & Accuracy & Precision & Recall & F1\\
\midrule
Weighted & 0.9595 & 0.9599 & 0.9627 & 0.9613\\
PSO & 0.9563 & 0.9723 & 0.9434 & 0.9577\\
Nelder--Mead & 0.9610 & 0.9627 & 0.9627 & 0.9627\\
Powell's method & 0.9532 & 0.9608 & 0.9494 & 0.9550\\
BFGS & 0.9532 & 0.9608 & 0.9494 & 0.9550\\
Trust-constr & 0.9532 & 0.9608 & 0.9494 & 0.9550\\
\bottomrule
\end{tabular}
\end{table}

\begin{table}[t]
\centering
\caption{Results of the large language models}
\label{tab:llm}
\begin{tabular}{lcccc}
\toprule
Model & Accuracy & Precision & Recall & F1\\
\midrule
GPT-2 & \textbf{0.9631} & \textbf{0.9612} & \textbf{0.9610} & \textbf{0.9600}\\
T5 & 0.9299 & 0.9356 & 0.9314 & 0.9310\\
XLNet & 0.9532 & 0.9540 & 0.9568 & 0.9554\\
\bottomrule
\end{tabular}
\end{table}

\begin{table}[t]
\centering
\caption{Comparison of state-of-the-art methods with the proposed work (accuracy)}
\label{tab:sota}
\begin{tabular}{ccc}
\toprule
\cite{b56} & \cite{b57} & Proposed (GPT-2)\\
\midrule
0.7900 & 0.6100 & \textbf{0.9631}\\
\bottomrule
\end{tabular}
\end{table}

\begin{figure}[t]
\centering
\begin{tikzpicture}
\begin{axis}[
  ybar, bar width=6pt, width=0.98\columnwidth, height=5cm,
  ymin=0.90, ymax=0.98, ylabel={Score},
  symbolic x coords={BERT-base,XLM-RoBERTa,ALBERT-base,DistilBERT}, xtick=data,
  tick label style={font=\scriptsize}, label style={font=\footnotesize},
  yticklabel style={/pgf/number format/fixed, /pgf/number format/precision=2},
  legend style={font=\scriptsize, at={(0.5,-0.2)}, anchor=north, legend columns=3},
  enlarge x limits=0.15]
\addplot[fill=blue!70] coordinates {(BERT-base,0.9540) (XLM-RoBERTa,0.9540) (ALBERT-base,0.9540) (DistilBERT,0.9485)};
\addplot[fill=orange!70] coordinates {(BERT-base,0.9568) (XLM-RoBERTa,0.9568) (ALBERT-base,0.9568) (DistilBERT,0.9598)};
\addplot[fill=teal!70] coordinates {(BERT-base,0.9554) (XLM-RoBERTa,0.9554) (ALBERT-base,0.9554) (DistilBERT,0.9541)};
\legend{Precision,Recall,F1-score}
\end{axis}
\end{tikzpicture}
\caption{Precision, recall, and F1-score of the TL models.}
\label{fig:tlmetrics}
\end{figure}
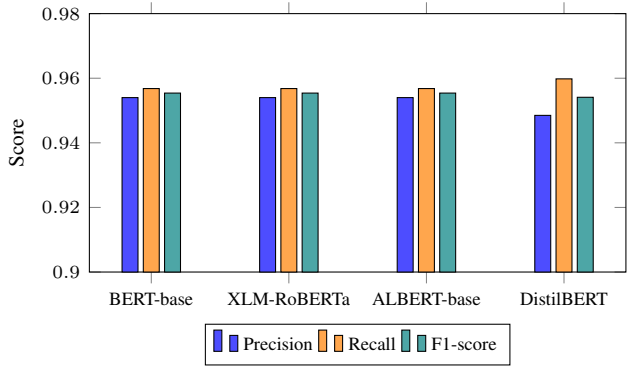

\section{Discussion}\label{sec:discussion}
Transformer-based models demonstrated remarkable performance on the COVID-19 English Tweets dataset. GPT-2 was identified as the best-performing model, achieving an accuracy of 96.31\%, with precision, recall, and F1-score of 96.12\%, 96.10\%, and 96.00\%, respectively. XLM-RoBERTa-base and ALBERT-base followed closely, each achieving an accuracy of 95.32\%. The results of all experiments show that the transformer-based models and LLMs outperformed the state-of-the-art machine learning approaches compared in Table~\ref{tab:sota}.

When several models produce similar results, this may be attributed to the use of a small and balanced dataset. Additionally, the use of similar hyperparameters or initialization methods may cause the models to converge toward similar solutions, resulting in nearly identical performance.

\section{Conclusion}\label{sec:conclusion}
In this study, four transfer learning models, namely BERT-base, XLM-RoBERTa-base, DistilBERT, and ALBERT-base, and three large language models, namely GPT-2, T5, and XLNet, were applied to the COVID-19 English Tweets dataset for binary text classification. The dataset was obtained from a reliable repository and contains real and fake English-language tweets. All models were evaluated using accuracy, recall, precision, and F1-score. Several optimization techniques were also explored, including weighted fusion, PSO fusion, Nelder--Mead fusion, Powell's method, BFGS fusion, and trust-region constrained fusion.

GPT-2 was identified as the best-performing model, achieving an accuracy of 0.9631, a precision of 0.9612, a recall of 0.9610, and an F1-score of 0.9600, demonstrating its effectiveness in classifying real and fake news.

In future work, we plan to conduct experiments using larger datasets for multiclass classification and to explore datasets in different languages. Additionally, we plan to use Twitter's live stream to collect real-time tweets and to investigate data from other social media platforms.


\begin{thebibliography}{99}
\bibitem{b1} Z. Zeng, H. Shi, Y. Wu, and Z. Hong, ``Survey of natural language processing techniques in bioinformatics,'' \emph{Computational and Mathematical Methods in Medicine}, vol. 2015, Article ID 674296, 2015.
\bibitem{b2} C. Jefferson, H. Liu, and M. Cocea, ``Fuzzy approach for sentiment analysis,'' in \emph{Proc. 2017 IEEE Int. Conf. Fuzzy Systems (FUZZ-IEEE)}, Naples, Italy, Jul. 2017.
\bibitem{b3} R. U. Mustafa \emph{et al.}, ``Early detection of controversial Urdu speeches from social media,'' \emph{Data Sci. Pattern Recognit.}, vol. 1, no. 2, pp. 26--42, 2017.
\bibitem{b4} K. Kumar, B. S. Harish, and H. K. Darshan, ``Sentiment analysis on IMDb movie reviews using hybrid feature extraction method,'' \emph{Int. J. Interactive Multimedia and Artificial Intelligence}, vol. 5, no. 5, p. 109, 2019.
\bibitem{b5} A. Schmidt and M. Wiegand, ``A survey on hate speech detection using natural language processing,'' in \emph{Proc. 5th Int. Workshop on Natural Language Processing for Social Media}, pp. 1--10, 2017.
\bibitem{b6} A. P. Kirilenko, S. O. Stepchenkova, H. Kim, and X. Li, ``Automated sentiment analysis in tourism: comparison of approaches,'' \emph{J. Travel Research}, vol. 57, no. 8, pp. 1012--1025, 2018.
\bibitem{b7} W. H. Bangyal, J. Ahmed, and H. T. Rauf, ``A modified bat algorithm with torus walk for solving global optimisation problems,'' \emph{Int. J. Bio-Inspired Computation}, vol. 15, no. 1, pp. 1--13, 2020.
\bibitem{b8} S. Pervaiz, Z. Ul-Qayyum, W. H. Bangyal, L. Gao, and J. Ahmad, ``A systematic literature review on particle swarm optimization techniques for medical diseases detection,'' \emph{Computational and Mathematical Methods in Medicine}, vol. 2021, Article ID 5990999, 2021.
\bibitem{b9} A. Ashraf, A. A. Almazroi, W. H. Bangyal, and M. A. Alqarni, ``Particle swarm optimization with new initializing technique to solve global optimization problems,'' \emph{Intelligent Automation \& Soft Computing}, vol. 31, no. 1, pp. 191--206, 2022.
\bibitem{b10} M. Mozafari, R. Farahbakhsh, and N. Crespi, ``A BERT-based transfer learning approach for hate speech detection in online social media,'' in \emph{Complex Networks and Their Applications VIII}, vol. 881, pp. 928--940, 2020.
\bibitem{b11} K. Florio, V. Basile, M. Polignano, P. Basile, and V. Patti, ``Time of your hate: the challenge of time in hate speech detection on social media,'' \emph{Applied Sciences}, vol. 10, no. 12, p. 4180, 2020.
\bibitem{b12} E. M. Okoro, B. A. Abara, A. O. Umagba, A. A. Ajonye, and Z. S. Isa, ``A hybrid approach to fake news detection on social media,'' \emph{Nigerian J. Technology}, vol. 37, no. 2, p. 454, 2018.
\bibitem{b13} A. Malte and P. Ratadiya, ``Evolution of transfer learning in natural language processing,'' arXiv:1910.07370, 2019.
\bibitem{b14} N. Houlsby \emph{et al.}, ``Parameter-efficient transfer learning for NLP,'' arXiv:1902.00751, 2019.
\bibitem{b15} A. Glazkova, M. Glazkov, and T. Trifonov, ``g2tmn at Constraint@AAAI2021: exploiting CT-BERT and ensembling learning for COVID-19 fake news detection,'' arXiv, 2020.
\bibitem{b16} S. Gundapu and R. Mamidi, ``Transformer based automatic COVID-19 fake news detection system,'' arXiv:2101.00180, 2021.
\bibitem{b17} S. Yang, K. Shu, S. Wang, R. Gu, F. Wu, and H. Liu, ``Unsupervised fake news detection on social media: a generative approach,'' in \emph{Proc. 33rd AAAI Conf. Artificial Intelligence}, pp. 5644--5651, Honolulu, HI, USA, 2019.
\bibitem{b18} A. Wani, I. Joshi, S. Khandve, V. Wagh, and R. Joshi, ``Evaluating deep learning approaches for Covid19 fake news detection,'' arXiv:2101.04012, 2021.
\bibitem{b19} T. Raha, V. Indurthi, and A. Upadhyaya, ``Identifying COVID-19 fake news in social media,'' arXiv:2101.11954, 2021.
\bibitem{b20} K. Chatsiou, ``Text classification of manifestos and COVID-19 press briefings using BERT and convolutional neural networks,'' arXiv, 2020.
\bibitem{b21} K. Mishev, A. Gjorgjevikj, I. Vodenska, L. T. Chitkushev, and D. Trajanov, ``Evaluation of sentiment analysis in finance: from lexicons to transformers,'' \emph{IEEE Access}, vol. 8, pp. 131662--131682, 2020.
\bibitem{b22} S. Yu, J. Su, and D. Luo, ``Improving BERT-based text classification with auxiliary sentence and domain knowledge,'' \emph{IEEE Access}, vol. 7, pp. 176600--176612, 2019.
\bibitem{b23} M.-A. Rizoiu, T. Wang, G. Ferraro, and H. Suominen, ``Transfer learning for hate speech detection in social media,'' arXiv, 2019.
\bibitem{b24} R. Qasim, W. H. Bangyal, M. A. Alqarni, and A. A. Almazroi, ``A fine-tuned BERT-based transfer learning approach for text classification,'' \emph{J. Healthcare Engineering}, vol. 2022, 2022.
\bibitem{b25} J. Devlin, M.-W. Chang, K. Lee, and K. Toutanova, ``BERT: pre-training of deep bidirectional transformers for language understanding,'' in \emph{Proc. NAACL-HLT}, pp. 4171--4186, 2019.
\bibitem{b26} R. K. Kaliyar, A. Goswami, and P. Narang, ``FakeBERT: fake news detection in social media with a BERT-based deep learning approach,'' \emph{Multimedia Tools and Applications}, vol. 80, no. 8, pp. 11765--11788, 2021.
\bibitem{b27} T. Guo and H. Gao, ``Content enhanced BERT-based text-to-SQL generation,'' arXiv, 2019.
\bibitem{b28} Z. Chen and T. Qian, ``Transfer capsule network for aspect level sentiment classification,'' in \emph{Proc. 57th Annual Meeting of the ACL}, pp. 547--556, Florence, Italy, 2019.
\bibitem{b29} C. Sun, L. Huang, and X. Qiu, ``Utilizing BERT for aspect-based sentiment analysis via constructing auxiliary sentence,'' in \emph{Proc. NAACL-HLT}, vol. 1, pp. 380--385, 2019.
\bibitem{b30} Z. Lan, M. Chen, S. Goodman, K. Gimpel, P. Sharma, and R. Soricut, ``ALBERT: a lite BERT for self-supervised learning of language representations,'' arXiv:1909.11942, 2019.
\bibitem{b31} H. Gregory, S. Li, P. Mohammadi, N. Tarn, R. Draelos, and C. Rudin, ``A transformer approach to contextual sarcasm detection in Twitter,'' in \emph{Proc. 2nd Workshop on Figurative Language Processing}, pp. 270--275, 2020.
\bibitem{b32} R. Vijjali, P. Potluri, S. Kumar, and S. Teki, ``Two stage transformer model for COVID-19 fake news detection and fact checking,'' arXiv:2011.13253, 2020.
\bibitem{b33} H. Ngai, Y. Park, J. Chen, and M. Parsapoor, ``Transformer-based models for question answering on COVID19,'' arXiv:2101.11432, 2021.
\bibitem{b34} M. A. Al-Garadi, Y.-C. Yang, H. Cai \emph{et al.}, ``Text classification models for the automatic detection of nonmedical prescription medication use from social media,'' \emph{BMC Medical Informatics and Decision Making}, vol. 21, no. 1, pp. 1--13, 2021.
\bibitem{b35} V. Sanh, L. Debut, J. Chaumond, and T. Wolf, ``DistilBERT, a distilled version of BERT: smaller, faster, cheaper and lighter,'' arXiv:1910.01108, 2019.
\bibitem{b36} B. B\"uy\"uk\"oz, A. H\"urriyeto\u{g}lu, and A. \"Ozg\"ur, ``Analyzing ELMo and DistilBERT on socio-political news classification,'' in \emph{Proc. Workshop on Automated Extraction of Socio-political Events from News 2020}, pp. 9--18, 2020.
\bibitem{b37} I. Staliunaite and I. Iacobacci, ``Compositional and lexical semantics in RoBERTa, BERT and DistilBERT: a case study on CoQA,'' in \emph{Proc. EMNLP}, pp. 7046--7056, 2020.
\bibitem{b38} M. Abadeer, ``Assessment of DistilBERT performance on named entity recognition task for the detection of protected health information and medical concepts,'' in \emph{Proc. 3rd Clinical Natural Language Processing Workshop}, pp. 158--167, 2020.
\bibitem{b39} J. Mozafari, A. Fatemi, and P. Moradi, ``A method for answer selection using DistilBERT and important words,'' in \emph{Proc. 6th Int. Conf. Web Research (ICWR)}, pp. 72--76, Tehran, Iran, 2020.
\bibitem{b40} P. Jayarao and A. Sharma, ``Retraining DistilBERT for a voice shopping assistant by using universal dependencies,'' arXiv:2103.15737, 2021.
\bibitem{b41} Y. Liu, M. Ott, N. Goyal \emph{et al.}, ``RoBERTa: a robustly optimized BERT pretraining approach,'' arXiv:1907.11692, 2019.
\bibitem{b42} A. Warstadt, Y. Zhang, X. Li, H. Liu, and S. R. Bowman, ``Learning which features matter: RoBERTa acquires a preference for linguistic generalizations (eventually),'' in \emph{Proc. EMNLP}, pp. 217--235, 2020.
\bibitem{b43} A. Murarka, B. Radhakrishnan, and S. Ravichandran, ``Detection and classification of mental illnesses on social media using RoBERTa,'' arXiv:2011.11226, 2020.
\bibitem{b44} S. Dhanalaxmi, R. Agarwal, and A. Sinha, ``Detection of COVID-19 informative tweets using RoBERTa,'' arXiv:2010.11238, 2020.
\bibitem{b45} Y. Zhao and X. Tao, ``ZYJ123@DravidianLangTech-EACL2021: offensive language identification based on XLM-RoBERTa with DPCNN,'' in \emph{Proc. 1st Workshop on Speech and Language Technologies for Dravidian Languages}, pp. 216--221, 2021.
\bibitem{b46} X. Ou and H. Li, ``YNU@Dravidian-CodeMix-FIRE2020: XLM-RoBERTa for multi-language sentiment analysis,'' pp. 4--9, 2020.
\bibitem{b47} X. Ou and H. Li, ``YNU\_OXZ @ HaSpeeDe 2 and AMI: XLM-RoBERTa with ordered neurons LSTM for classification task at EVALITA 2020,'' in \emph{EVALITA 2020}, CEUR vol. 2765, pp. 102--109, 2020.
\bibitem{b48} S. Xie, J. Ma, H. Yang, L. Jiang, Y. Mo, and J. Shen, ``PALI at SemEval-2021 Task 2: fine-tune XLM-RoBERTa for word in context disambiguation,'' arXiv:2104.10375, 2021.
\bibitem{b49} S. S. Tekiroglu, H. Bonaldi, M. Fanton, and M. Guerini, ``Using pre-trained language models for producing counter narratives against hate speech: a comparative study,'' arXiv:2204.01440, 2022.
\bibitem{b50} M. Tayyab Zamir, M. A. Ayub, A. Gul, N. Ahmad, and K. Ahmad, ``Stylometry analysis of multi-authored documents for authorship and author style change detection,'' arXiv preprint, 2024.
\bibitem{b51} K. Ahmad, M. A. Ayub, J. Khan, N. Ahmad, and A. Al-Fuqaha, ``Social media as an instant source of feedback on water quality,'' \emph{IEEE Trans. Technology and Society}, 2022.
\bibitem{b52} J. Kennedy and R. Eberhart, ``Particle swarm optimization,'' in \emph{Proc. ICNN'95 Int. Conf. Neural Networks}, vol. 4, pp. 1942--1948, 1995.
\bibitem{b53} B. Qolomany, K. Ahmad, A. Al-Fuqaha, and J. Qadir, ``Particle swarm optimized federated learning for industrial IoT and smart city services,'' in \emph{Proc. IEEE GLOBECOM}, pp. 1--6, 2020.
\bibitem{b54} Y. Wang, H. Zhang, and G. Zhang, ``cPSO-CNN: an efficient PSO-based algorithm for fine-tuning hyper-parameters of convolutional neural networks,'' \emph{Swarm and Evolutionary Computation}, vol. 49, pp. 114--123, 2019.
\bibitem{b55} S. Ohashi, T. Kajiwara, C. Chu, N. Takemura, Y. Nakashima, and H. Nagahara, ``IDSOU at WNUT-2020 Task 2: identification of informative COVID-19 English tweets,'' in \emph{Proc. 6th Workshop on Noisy User-generated Text (W-NUT 2020)}, pp. 428--433, 2020.
\bibitem{b56} C. Monica and N. Nagarathna, ``Detection of fake tweets using sentiment analysis,'' \emph{SN Computer Science}, vol. 1, no. 2, p. 89, 2020.
\bibitem{b57} A. Khalil, H. Hajjdiab, and N. Al-Qirim, ``Detecting fake followers in Twitter: a machine learning approach,'' \emph{Int. J. Machine Learning and Computing}, vol. 7, no. 6, pp. 198--202, 2017.
\bibitem{b58} M. O. Topal, A. Bas, and I. van Heerden, ``Exploring transformers in natural language generation: GPT, BERT, and XLNet,'' arXiv:2102.08036, 2021.
\bibitem{b59} J. Ni, G. H. Abrego, N. Constant, J. Ma, K. B. Hall, D. Cer, and Y. Yang, ``Sentence-T5: scalable sentence encoders from pre-trained text-to-text models,'' arXiv:2108.08877, 2021.
\end{thebibliography}
\end{document}